\documentclass[lettersize,journal]{IEEEtran}
\usepackage{amsmath,amsfonts}
\usepackage{algorithmic}
\usepackage{array}
\usepackage[caption=false,font=normalsize,labelfont=sf,textfont=sf]{subfig}
\usepackage{textcomp}
\usepackage{stfloats}
\usepackage{url}
\usepackage{verbatim}
\usepackage{graphicx}
\def\BibTeX{{\rm B\kern-.05em{\sc i\kern-.025em b}\kern-.08em
    T\kern-.1667em\lower.7ex\hbox{E}\kern-.125emX}}
\usepackage{balance}
\usepackage{bm}
\usepackage{booktabs}
\usepackage{multirow}
\usepackage{upgreek} 

\begin{document}
\title{Acceleration of Multi-Scale LTS Magnet Simulations with Neural Network Surrogate Models}
\author{Louis Denis, Julien Dular, Vincent Nuttens, Mariusz Wozniak, Benoît Vanderheyden, and Christophe Geuzaine
\thanks{L. Denis is a research fellow of the Fonds de la Recherche Scientifique - FNRS. The work of J. Dular and M. Wozniak is partially supported by the CERN HFM program. (\textit{Corresponding author: L. Denis})

L. Denis, B. Vanderheyden, and C. Geuzaine are with the Department of Electrical Engineering and Computer Science, Institut Montefiore B28 in the University of Liege, 4000 Liege, Belgium (e-mail: louis.denis@uliege.be).

J. Dular and M. Wozniak are with CERN, 1211 Meyrin, Switzerland.

V. Nuttens is with Ion Beam Applications, 1348 Louvain-la-Neuve, Belgium.}}

\markboth{September 2025}
{L. Denis, J. Dular, V. Nuttens, M. Wozniak, B. Vanderheyden, and C. Geuzaine: Acceleration of Multi-Scale LTS Magnet Simulations with Neural Network Surrogate Models}

\makeatletter
\def\ps@IEEEtitlepagestyle{
  \def\@oddfoot{\mycopyrightnotice}
  \def\@evenfoot{}
}
\def\mycopyrightnotice{
  {\footnotesize
  \begin{minipage}{\textwidth}
    \fbox{\parbox{\textwidth}{
        This work has been submitted to a journal for possible publication. Copyright may be transferred without notice, after which this version may no longer be accessible.
        }}
  \end{minipage}
  }
}

\maketitle

\begin{abstract}
While the prediction of AC losses during transients is critical for designing large-scale low-temperature superconducting (LTS) magnets, brute-force finite-element (FE) simulation of their detailed geometry down to the length scale of the conductors is a computational challenge. Multi-scale methods, balancing between a coarse approximation of the fields at the scale of the magnet and a detailed description at the scale of the conductors, are promising approaches to reduce the computational load while keeping a sufficient accuracy.

In this work, we introduce a neural network approach to accelerate multi-scale magneto-thermal simulations of LTS magnets by replacing costly single-turn FE models with neural network surrogates. The neural network architecture is presented and discussed, together with an automated procedure for generating simulation data for its training. The resulting accelerated multi-scale model is used to simulate current ramp-up procedures for the IBA S2C2 magnet. The surrogate-based multi-scale model is compared with a conventional multi-scale model based on a composite wire-in-channel FE model. The surrogate model is shown to reproduce single-turn filament hysteresis, inter-filament coupling, and eddy losses, while the computational time of the multi-scale method is reduced by a factor of 800.
\end{abstract}


\section{Introduction}
\IEEEPARstart{D}{esigning} large-scale Low-Temperature Superconducting (LTS) magnets for industrial applications requires fast and reliable AC loss simulations. Because the main loss mechanisms (filament hysteresis, inter-filament coupling and eddy losses) occur at the scale of individual conductors~\cite{Campbell1982, Wilson1983, Dular2025ah}, an accurate estimation of the AC losses involves the calculation of the magneto-thermal response of the system from the scale of single conductors to the macroscopic scale of the magnet. While recent parallelization methods, such as domain decomposition~\cite{Riva2023} or GPU-optimized algorithms~\cite{Xue2024}, can accelerate resolutions, full-scale finite-element (FE) simulations are currently out of reach for desktop computers.

In this context, computational homogenization~\cite{ElFeddi1997} via multi-scale resolution techniques~\cite{Feyel2003, Abdulle2009} can be employed to reduce simulation times. The main idea is to couple a macroscopic FE model describing averaged physical fields with local mesoscopic-scale models based on a detailed geometry, in order to capture, e.g., the relationship between local loss mechanisms and the large-scale temperature distribution in LTS magnets. In a recent work~\cite{Denis2025}, we introduced a magneto-thermal multi-scale approach (MSA) that combined an homogenized 3D FE LTS magnet model with meso-scale models estimating AC local losses at the level of single filaments (SF).

In this work, the MSA is first improved by modelling the actual composite superconductor geometry at the mesoscopic scale, replacing the SF model in~\cite{Denis2025} and enabling the representation of the different loss mechanisms with the FE method. While twisted superconductors are here modelled efficiently using the CATI method~\cite{Dular2024}, their simulation in a full multi-scale resolution remains computationally expensive as it still requires high-performance computing facilities. In the present work, substantial acceleration of multi-scale simulations is achieved by replacing the mesoscopic-scale FE model with a neural network (NN) surrogate, an approach that showed promising results in computational mechanics~\cite{Wu2020}. Such NN surrogates represent an alternative to classical model order reduction techniques, such as energy-based models~\cite{Bergqvist1997,Henrotte2006} recently adapted to superconductors~\cite{Dular2025rohm,Glock2025}.

The use of machine learning-based surrogates in applied superconductivity has grown rapidly in recent years~\cite{Tomassetti2025} and offers interesting perspectives in modelling and optimizing superconducting systems~\cite{YazdaniAsrami2023}. Among others, physics-informed NNs are being investigated as alternatives to complete FE solvers~\cite{Raissi2019,AlipourBonab2025}. Here, we are specifically interested in the time-evolution of AC losses in composite superconductors. To this end, we opt for a NN surrogate based on Gated Recurrent Units (GRUs)~\cite{Cho2014} designed to represent history-dependent phenomena such as hysteresis in superconductors. This approach differs from previous NN surrogate for AC losses in single conductors~\cite{Leclerc2016,YazdaniAsrami2020,Zhou2025}, mostly designed for periodic excitations and limited to averaged loss predictions. Moreover, our approach learns the direct mapping between time sequences of fields, avoiding the intermediate step of fitting coefficients in a parametric constitutive law~\cite{Purnode2024,Dular2025rohm}.

Section II introduces the generalized FE-based MSA and its GRU-based counterpart. Section III discusses the NN surrogate, its architecture, its automated dataset generation and training procedure. Both the FE-based MSA and the novel GRU-based MSA are applied to ramp-up simulations of the IBA S2C2 synchrocyclotron~\cite{Kleeven2013} in Section IV, demonstrating the significant speedup achieved with the NN surrogate.

\section{Multi-scale Approach}
\begin{figure}[!t]
\centering
\includegraphics[scale=0.9]{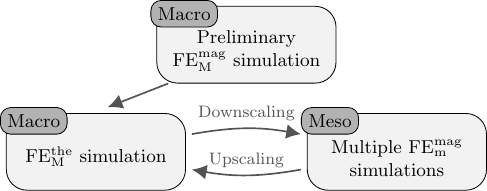}
\caption{Flowchart of the FE-based MSA, coupling the macroscopic thermal FE simulation of the LTS coil with mesoscopic magnetodynamic FE simulations.}
\label{fig:msa-principle}
\end{figure}
The basic scheme of the FE-based multi-scale approach is detailed in~\cite{Denis2025}, and its generic workflow is shown in Fig.~\ref{fig:msa-principle}. Subscripts $\cdot_{\text{M}}$ and $\cdot_{\text{m}}$ denote macro- and mesoscopic-scale fields, respectively. In the LTS magnet, conductor turns are assumed to be connected in series and with negligible transverse dimensions w.r.t. the magnet radius, so that the coupling between neighbouring individual turns is neglected at the macroscopic scale. Given a prescribed current source $I_{\text{M}}(t)$, the macroscopic magnetic flux density $\bm{b}_{\text{M}}(\bm{x}_{\text{M}},t)$ in the magnet is thus obtained from a preliminary magnetodynamic simulation FE$_\text{M}^{\text{mag}}$ (e.g. with the $a$-$v$ formulation as in~\cite{Denis2025}).

The second step of the MSA involves using a macroscopic thermal simulation FE$_\text{M}^{\text{the}}$ to solve the heat equation for the temperature field $T_{\text{M}}$ within the cold mass of the magnet:
\begin{equation}
\rho_{T} c_p \partial_t T_{\text{M}} + \text{div} \cdot (-\bm{\kappa} \cdot \textbf{grad}~T_{\text{M}}) = q_{\text{s}}, \label{eq:heat_eq}
\end{equation}
where $\rho_{T}$, $c_p$ and $\bm{\kappa}$ are thermal material properties that have been homogenized (cf.~\cite[Appendix A]{Denis2025}) and $q_{\text{s}}$ is the heat source (in W/m$^3$). The combination of~\eqref{eq:heat_eq} with spatial boundary conditions (e.g. cooling via cryocoolers) typically requires 3D simulations.

In normal conductors (e.g. copper bars, aluminum coil formers), the heat source corresponds to conventional Joule losses $q_{\text{s}} = \rho_{\text{M}} \lVert\bm{j}_{\text{M}} \rVert^2$, with $\rho_{\text{M}}$ the electrical resistivity and $\bm{j}_{\text{M}}$ the current density from the macroscopic simulation. In composite superconductors (i.e. in the LTS magnet), the loss density can be decomposed into its three main contributions:
\begin{equation}
    q_{\text{s}} = q_{\text{fil}} + q_{\text{coupl}} + q_{\text{eddy}}, \label{eq:loss_decomposition}
\end{equation}
with $q_{\text{fil}}$, $q_{\text{coupl}}$ and $q_{\text{eddy}}$ denoting filament hysteresis, inter-filament coupling and eddy losses~\cite{Dular2024}, respectively. While some of these components can be estimated analytically \cite{Campbell1982, Wilson1983, Carr2001,Denis2025}, the FE-based MSA relies on multiple mesoscopic-scale models FE$_\text{m}^{\text{mag}}$ of a single conductor turn for reliable loss prediction. In particular, the MSA in~\cite{Denis2025} is generalized here by replacing its meso-scale single-filament model with a more comprehensive model of the composite superconductor. The mesoscopic FE model determines the averaged loss components in \eqref{eq:loss_decomposition} by solving for the local magnetic field $\bm{h}_{\text{m}}$ and its corresponding current density $\bm{j}_{\text{m}}$, given transport current $I_{\text{M}}(t)$ and applied external flux density $\bm{b}_{\text{M}}(t)$ sequences from the macroscopic scale. Due to the high thermal conductivity of copper ensuring rapid thermal equilibrium, temperature is not solved at the mesoscopic scale and is assumed uniform $T_{\text{m}} = T_{\text{M}}$ within each turn. Still, mesoscopic losses are dependent on $T_{\text{M}}$ via the local critical current density $j_{\text{c}}(\bm{b}_{\text{m}},T_{\text{M}})$. 

\begin{figure}[!t]
\centering
\includegraphics[width=\columnwidth]{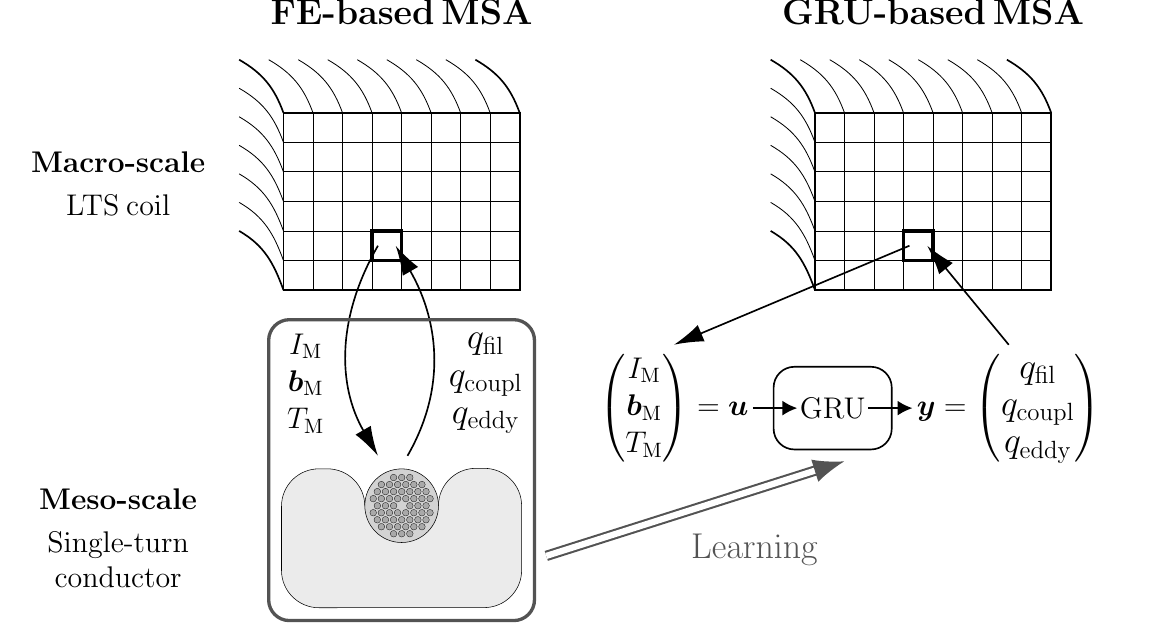}
\caption{FE-based MSA: fields are downscaled from the macroscopic LTS coil model to the mesoscopic conductor model and AC losses are upscaled back (left). GRU-based MSA: the mesoscopic model FE$_\text{m}^{\text{mag}}$ is replaced with a GRU-based NN surrogate (right).}
\label{fig:detailed-msa}
\end{figure}

Consequently, the FE-based MSA relies on a two-way coupling between the macroscopic thermal simulation FE$_\text{M}^{\text{the}}$ and multiple mesoscopic loss simulations FE$_\text{m}^{\text{mag}}$. This is illustrated in Fig.~\ref{fig:detailed-msa} for a wire-in-channel (WIC) conductor, as used in the IBA S2C2 magnet~\cite{Denis2025}. Rather than associating one mesoscopic model FE$_\text{m}^{\text{mag}}$ with each macroscopic integration point, as in the FE$^2$ method~\cite{Feyel2003}, the coil cross-section is subdivided into zones and the FE-based MSA in Fig.~\ref{fig:detailed-msa} assigns one mesoscopic model to each macroscopic zone. This improves efficiency as it greatly reduces the number of mesoscopic simulations required~\cite{Denis2025}. In this case, the inputs $\bm{u} = (I_{\text{M}}, \bm{b}_{\text{M}}, T_{\text{M}})^T$ of each single-turn model correspond to field sequences spatially averaged over the corresponding zone.

From the macroscopic perspective, the mesoscopic FE model acts as a black box returning the loss density contributions, that can be expressed at a given time $t^n = n \Delta t_{\text{M}}$ (with $\Delta t_{\text{M}}$ the time step of FE$_\text{M}^{\text{the}}$) as
\begin{equation}
    \bm{y}(t^n) = \bm{y}(\underbrace{\bm{u}(t \le t^n)}_{\text{history}},\text{conductor properties, geometry}), \label{eq:history}
\end{equation}
where $\bm{y} = (q_{\text{fil}},q_{\text{coupl}},q_{\text{eddy}})^T$. Since the conductor features are fixed for a given magnet, \eqref{eq:history} shows that a history-dependent surrogate could replace the single-turn FE model for faster loss prediction. A GRU-based NN surrogate is considered for this purpose and its integration in the MSA is represented in Fig.~\ref{fig:detailed-msa}. The GRU-based MSA is thus similar to the FE-based MSA, with the mesoscopic FE model (bottom left in Fig.~\ref{fig:msa-principle}) replaced by the surrogate. Both FE-based and GRU-based MSAs are implemented in GetDP~\cite{Dular1998} with a Python interface for communication between scales.

\section{Neural Network Surrogate}
\subsection{Neural Network Architecture}
\begin{figure}[!t]
\centering
    \includegraphics[width=\columnwidth]{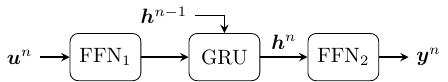}
    \caption{Structure of the GRU-based RNN. Two feedforward networks FFNs are placed in front and after the GRU, respectively.}
    \label{fig:GRU_architecture}
\end{figure}

GRUs are designed to capture temporal dependencies by treating data sequentially, with the history at time $t_n$ encoded in the hidden state vector $\bm{h}^{n}$~\cite{Cho2014}. The simplified GRU-based surrogate architecture is shown in Fig.~\ref{fig:GRU_architecture}. At time $t_n$, the input $\bm{u}^{n}$ is first processed by a feedforward network (FFN$_1$), whose output is passed to the GRU along with its previous hidden state $\bm{h}^{n-1}$. The GRU (one hidden layer of size 100, initialized to zero) produces the updated $\bm{h}^{n}$, which is then passed through a second feedforward network (FFN$_2$) to generate the final predictions~$\bm{y}^{n}$. FFN$_1$ and FFN$_2$ each have one hidden layer of size 32, with LeakyReLU and Sigmoid activation functions, respectively. The optimal hyperparameter values were determined via trial and error.

Enriching the input vector $\bm{u}$ with additional features increased accuracy. The optimal setting includes eight scaled macroscopic features, denoted $\Tilde\cdot$,
\begin{equation}
    \bm{u}^{n} = \left(\Tilde I_{\text{M}}^n, \Tilde{b}_{\text{M},r}^n, \Tilde{b}_{\text{M},z}^n, \Tilde T_{\text{M}}^{n-1}, \Tilde{t}^n, \partial_t \Tilde I_{\text{M}}^n, \partial_t \Tilde{b}_{\text{M},r}^n, \partial_t \Tilde{b}_{\text{M},z}^n\right)^T, \label{eq:input-features}
\end{equation}
where $b_{\text{M},r}$ and $b_{\text{M},z}$ are the radial and vertical components of the macroscopic flux density. In particular, passing the macroscopic temperature at the previous time step $ T_{\text{M}}^{n-1}$ was found to increase the efficiency of the GRU-based MSA.

For a current ramp-up of the LTS magnet, input features $u_i \in [u_i^{\text{min}},u_i^{\text{max}}]$ are linearly scaled to $\Tilde u_i \in [-1,1]$ for ${b_{\text{M},r}, b_{\text{M},z}, \partial_t b_{\text{M},r}, \partial_t b_{\text{M},z}}$, and to $\Tilde u_i \in [0,1]$ for ${I_{\text{M}}, T_{\text{M}}, t, \partial_t I_{\text{M}}}$, where $u_i^{\text{min}}$ and $u_i^{\text{max}}$ are the minimal and maximal values in the dataset. To maintain accuracy across the full range of losses, each output feature $y_i$ is scaled as $\Tilde y_i = (y_i / y_i^{\text{max}})^{1/3}$. Combined with the Sigmoid final activation of the NN, this scaling provides a physical safeguard: predictions cannot exceed the loss values observed during training.

\subsection{Automated Training Procedure}
\begin{figure}[!t]
\centering
    \includegraphics[scale=0.75]{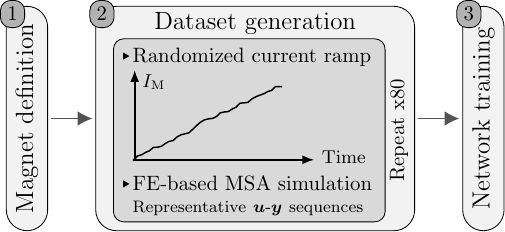}
    \caption{Three steps of the offline automated training procedure.}
    \label{fig:automated-training}
\end{figure}

A generic three-step automated training procedure is proposed in Fig.~\ref{fig:automated-training}, such that a surrogate model can be trained for each particular LTS magnet to be studied. The first step requires human input. It consists in describing the macroscopic magnet model (geometry, material properties and FE solver), and the corresponding mesoscopic single conductor turn model. A sampling method (e.g. subdividing the coil cross-section in 100 zones) must be defined to retrieve representative field sequences.

The second step focuses on dataset generation. During magnet ramp-up, the control variable is the imposed current~$I_{\text{M}}$. 80 different current sequences are generated as random walks, with the constraint that current can only increase and must reach its nominal value after ramp-up: $I_{\text{M}}(T_{\text{up}})=I_{\text{nom}}$. One example sequence is shown in Fig.~\ref{fig:automated-training}. The total ramp-up time~$T_{\text{up}}$ is varied to cover different time scales. For the S2C2 magnet, it is drawn randomly in $[3600, 36000]$ (in s). This randomization of current profiles ensures a broad coverage of possible trajectories in both the input and output spaces, improving the surrogate’s generalization capacity. For each current profile, a FE-based MSA simulation is then run (cf. left part in Fig~\ref{fig:detailed-msa}). Each run produces a given amount of representative $\bm{u}$-$\bm{y}$ sequences (e.g. 1 per zone, thus 100 per run). Together, these sequences form the dataset.

The final step consists in training the NN surrogate using a subset of the generated dataset: 85\% of the sequences are used for training and 15\% for validation. Training is performed with PyTorch, the Adam optimizer and the mean squared error (MSE) loss function. Multiple training runs are executed with different learning parameters, and the model associated to the lowest validation MSE is selected.

\section{Application, Results and Discussion}
The GRU-based MSA is compared to the FE-based MSA as they are applied to the magnetothermal simulation of the S2C2 LTS magnet~\cite{Kleeven2013}. Its 3D macroscopic thermal model and its preliminary axisymmetric magnetic model, including the ferromagnetic yoke, are detailed in~\cite[Sec. VII.a]{Denis2025}. In the FE-based MSA, the WIC-geometry is assumed periodic and is solved using the CATI method~\cite{Dular2024} implemented in FiQuS~\cite{Vitrano2023}, which replaces the full 3D twisted conductor model with two coupled 2D finite-element problems. The WIC model is similar to the one described in~\cite[Sec. 5.2]{Dular2024} and sketched in Fig.~\ref{fig:detailed-msa}, with its main parameters adapted to the S2C2 conductor and gathered in Table~\ref{tab:wic-parameters}.

\begin{table}[t!]
    \vspace*{-.5em}
    \begin{center}
    \caption{Main parameters of the WIC conductor in the S2C2 coil.}
    \label{tab:wic-parameters}
    \vspace*{-0.5em}
    \begin{tabular}{cc}
    \toprule
    Description & Value \\
    \midrule
    Critical current density $j_{\text{c}}(b,T)$-fit & \cite{Bottura2000,Spencer1979,Lubell1983} \\
    $j_{\text{c}}(5~\text{T}, 4.2~\text{K})$ & $2783$~A/mm$^2$~\cite{Denis2025}\\
    Power-law $n$-index & $50$\\
    NbTi Filament diameter & $156$~$\upmu$m\\
    Superconductor filling ratio in WIC & $0.148$ \\
    Central strand twist pitch length & $0.1$~m \\
    Copper (strand matrix and channel) RRR & $80$\\
    Nominal current $I_{\text{nom}}$ & $656$~A\\
    \bottomrule
    \end{tabular}
    \end{center}
    \vspace*{-1em}
\end{table}

\begin{figure}[!t]
\centering
    \includegraphics{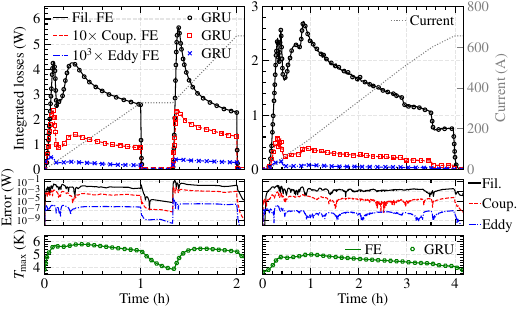}
    \caption{Volume-integrated filament hysteresis, coupling and eddy losses in the LTS coil, computed with FE-based and GRU-based MSAs (top), together with the corresponding absolute error (center) and maximal coil temperature (bottom). Example 2~h ramp-up composed of two linear ramps with an intermediate pause (left). Practical 4~h ramp-up profile of the S2C2~\cite{Denis2025} (right).}
    \label{fig:curves-comparison}
\end{figure}

The NN surrogate is trained by following the procedure described in Section III.B, and 80 FE-based MSA simulations are run with the LTS coil subdivided into 100 zones. With the same zone subdivision, the GRU-based MSA associates one NN surrogate to each zone. As shown in Fig.~\ref{fig:curves-comparison}, the GRU-based MSA reproduces FE-based MSA loss predictions when tested on practical engineering ramp-ups of different durations, including the practical current IBA S2C2 ramp-up profile. The current profiles in Fig.~\ref{fig:curves-comparison} are not part of the training dataset, highlighting the strong generalization capacity of the NN surrogate. The surrogate extracts the relevant information from the 49k degrees of freedom of the FE WIC model into 100 hidden states, without explicitly resolving the detailed current density $\bm{j}_{\text{m}}$ distribution. Notably, loss contributions of different physical nature, see~\eqref{eq:loss_decomposition}, and spanning several orders of magnitude are accurately captured. This is quantified by computing the coefficient of determination $R^2$ for each loss contribution $Q_{\text{GRU}}(t)$ predicted by the GRU-based MSA w.r.t. the FE-based reference $Q_{\text{FE}}(t)$ (mean value: $\bar{Q}_{\text{FE}}$), defined as
\begin{equation}
    R^2 = 1 - \frac{\int_{0}^{T_{\text{up}}} (Q_{\text{FE}}(t) - Q_{\text{GRU}}(t))^2\,dt}{\int_{0}^{T_{\text{up}}} (Q_{\text{FE}}(t) - \bar{Q}_{\text{FE}})^2\,dt}.
\end{equation}
The $1-R^2$ values gathered in Table~\ref{tab:r2-global-losses} confirm the excellent agreement between the GRU-based MSA and its FE-based counterpart, and the ability of the NN surrogate to produce reliable predictions in realistic operation conditions. As shown in Fig.~\ref{fig:curves-comparison}, the coil maximum temperature evolution is also well captured, with corresponding $1-R^2$ values of $1.23\times10^{-4}$ and $1.6\times10^{-5}$ for the 2~h and 4~h ramp-ups, respectively.

\begin{table}[t!]
    \vspace*{-1em}
    \begin{center}
    \caption{$1 - R^2$ coefficient of volume-integrated losses for the two illustrated current ramp-up profiles.}
    \label{tab:r2-global-losses}
    \vspace*{-0.5em}
    \begin{tabular}{c|ccc}
    \toprule
    Loss contribution & Filament & Coupling & Eddy \\
    \midrule
    Example 2~h ramp & $2.27 \times 10^{-4}$ & $2.93 \times 10^{-4}$ & $4.71 \times 10^{-4}$ \\
    Practical 4~h ramp & $2.57 \times 10^{-5}$ & $7.37 \times 10^{-5}$ & $3.68 \times 10^{-5}$ \\
    \bottomrule
    \end{tabular}
    \end{center}
    \vspace*{-2em}
\end{table}

The computational performance of the different approaches for the 4~h S2C2 ramp-up simulation is summarized in Table~\ref{tab:computational-cost-MSA}. A single FE-based MSA simulation requires 1500~CPU-h. In contrast, after training, the GRU-based MSA simulation (with one surrogate per macroscopic zone, or 100 in total) runs in less than 2~h on a single CPU, achieving an 800-fold speedup. Its memory requirement (in terms of the maximum resident set size RSS) is also significantly lower. The remarkable efficiency of the GRU-based MSA enables the evaluation of one distinct mesoscopic-scale surrogate at each of the 41k integration points in the FE$^{\text{the}}_{\text{M}}$ model, leading to a FE$^2$-type method. This is a significant step towards full-scale simulations of LTS magnets, since the FE-based MSA with as many mesoscopic models would require $41\text{k}\times15\approx600$k~CPU-h. The $1-R^2$ coefficients between the full 41k-points GRU-based MSA and the 100-zone GRU-based MSA remain below $5\times10^{-4}$, confirming that the zone-discretized approach is a reasonable approximation.

\begin{table}[t!]
    \begin{center}
    \caption{Computational performance figures.}
    \label{tab:computational-cost-MSA}
    \vspace*{-0.5em}
    \begin{tabular}{c|ccc}
    \toprule
    MSA & FE-based & GRU-based & GRU-based \\
    Nb. mesoscopic models & $100$ & $100$ & $41$k \\
    \midrule
    Wall time (s) & $54055$ & $6056$ & $7278$ \\
    Nb. CPUs & $100$ & $1$ & $1$ \\
    CPU hours & $1500$ & $1.7$ & $2.02$ \\
    Maximum RSS (GB) & $155$ & $0.664$ & $4.1$ \\ 
    \midrule
    Offline dataset generation & - & \multicolumn{2}{c}{$80 \times 1500=120$k~CPU-h} \\
    \bottomrule
    \end{tabular}
    \end{center}
    \vspace*{-1.5em}
\end{table}

Table~\ref{tab:computational-cost-MSA} also shows the significant cost of the training procedure. While the training of the NN itself takes less than one hour on a single GPU, the 80 FE-based MSA simulations required for dataset generation correspond to about 120k~CPU-h. This cost cannot be overlooked, and the proposed training strategy is justified only when a large number of simulations of a given magnet are needed. Among other applications, this approach is suited to the optimization of the magnet ramp-up profile, as will be investigated in future work.

\begin{figure}[!t]
\centering
    \includegraphics{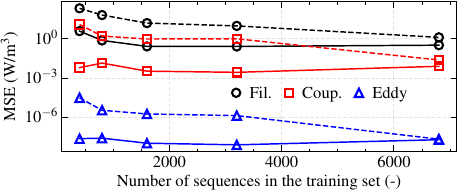}
    \caption{Training (solid lines) and validation (dashed lines) MSE of the best model (selected at minimal validation MSE) for each loss contribution, as a function of training set size. Each point represents the average over five independent runs to account for stochastic variability.}
    \label{fig:training-set-size-dependence}
    \vspace*{-.5em}
\end{figure}

The importance of generating sufficiently many input–output sequences is illustrated in Fig.~\ref{fig:training-set-size-dependence}, where each loss component was scaled back to its physical range. With a small training set, the NN surrogate overfits: the MSE is significantly lower when evaluated on the training set. As the training size increases, the generalization capacity of the NN improves and the gap between training and validation MSE decreases.

\section{Conclusion}
Multi-scale magneto-thermal FE simulations of LTS magnets have been accelerated by replacing costly mesoscopic FE models with a neural network surrogate. An automated framework for training the surrogate has been developed and applied to the 3D simulation of the IBA S2C2 magnet describing a current ramp-up. Once trained on detailed simulation data, the surrogate delivers an 800-fold speedup of the 3D multi-scale model while showing excellent agreement in predicting AC loss contributions of different physical origins. This approach provides a reliable tool for design, optimization, and real-time simulation of large-scale LTS magnets.

\section*{Acknowledgments}
The authors thank F. Henrotte, F. Purnode and D. Colignon (University of Liège) for valuable discussions on the implementation of the methods. The present research benefited from computational resources made available on Lucia, the Tier-1 supercomputer of the Walloon Region, infrastructure funded by the Walloon Region under the grant agreement n°1910247.

\balance


\begin{thebibliography}{99}

\bibitem{Campbell1982}
A. Campbell, “A general treatment of losses in multifilamentary superconductors,” \emph{Cryogenics}, vol. 22, no. 1, pp. 3-16, 1982.

\bibitem{Wilson1983}
M. N. Wilson, “Superconducting Magnets,” Oxford, U.K.: Clarendon, 1983.

\bibitem{Dular2025ah}
J. Dular, A. Verweij and M. Wozniak, “A Finite Element  a-h-Formulation for the Reduced Order Hysteretic Magnetization Model for Composite Superconductors,” \emph{IEEE Trans. Appl. Supercond.}, vol. 35, no. 5, Aug. 2025, Art no. 8200205.

\bibitem{Riva2023}
N. Riva \emph{et al.}, “$H $-$\phi$ Formulation in Sparselizard Combined With Domain Decomposition Methods for Modeling Superconducting Tapes, Stacks, and Twisted Wires,” \emph{IEEE Trans. Appl. Supercond.}, vol. 33, no. 5, Art. no. 4900405, 2023.

\bibitem{Xue2024}
C. Xue \emph{et al.}, “Holistic numerical simulation of a quenching process on a real-size multifilamentary superconducting coil,” \emph{Nature Communications}, vol. 15, no. 1, p. 10454, 2024.

\bibitem{ElFeddi1997}
M. El Feddi, Z. Ren, A. Razek, and A. Bossavit, “Homogenization technique for Maxwell equations in periodic structures,” \emph{IEEE Trans. Magn.}, vol. 33, no. 2, pp. 1382–1385, Mar. 1997.

\bibitem{Feyel2003}
F. Feyel, “A multilevel finite element method (FE2) to describe the response of highly non-linear structures using generalized continua", \emph{Comput. Methods Appl. Mech. Eng.,} vol. 192, no. 28-30, pp. 3233-3244, 2003.

\bibitem{Abdulle2009}
A. Abdulle, “The finite element heterogeneous multiscale method: a computational strategy for multiscale PDEs,” \emph{GAKUTO Int. Ser. Math. Sci. Appl.}, vol. 31, p. 135-184, 2009.

\bibitem{Denis2025}
L. Denis, V. Nuttens, B. Vanderheyden, and C. Geuzaine, “AC Loss Computation in Large-Scale Low-Temperature Superconducting Magnets: Multiscale and Semianalytical Procedures,” \emph{IEEE Trans. Appl. Supercond.}, vol. 35, no. 2, 2025, Art no. 5900319.

\bibitem{Dular2024}
J. Dular \emph{et al.}, “Coupled axial and transverse currents method for finite element modelling of periodic superconductors,” \emph{Supercond. Sci. Technol.}, vol. 37, 2024.

\bibitem{Wu2020}
L. Wu, N. G. Kilingar, and L. Noels, “A recurrent neural network-accelerated multi-scale model for elasto-plastic heterogeneous materials subjected to random cyclic and non-proportional loading paths,” \emph{Comput. Meth. Appl. Mech. Eng.}, vol. 369, 2020.

\bibitem{Bergqvist1997}
A. Bergqvist, “Magnetic vector hysteresis model with dry friction-like pinning,” \emph{Physica B: Condensed Matter}, vol. 233, no. 4, pp. 342-347, 1997.

\bibitem{Henrotte2006}
F. Henrotte, A. Nicolet, and K. Hameyer, “An energy-based vector hysteresis model for ferromagnetic materials,” \emph{COMPEL: The International Journal for Computation and Mathematics in Electrical and Electronic Engineering}, vol. 25, no. 1, pp. 71-80, 2006.

\bibitem{Dular2025rohm}
J.Dular, A. Verweij, and M. Wozniak, “Reduced Order Hysteretic Magnetization Model for Composite Superconductors,” \emph{Supercond. Sci. Technol.}, vol. 38, no. 3, p. 035017, 2025.

\bibitem{Glock2025}
A. Glock, J. Dular, A. Verweij, and M. Wozniak, “Reduced Order Hysteretic Flux Model for Transport Current Homogenization in Composite Superconductors,” \emph{arXiv:2507.15402}, 2025.

\bibitem{Tomassetti2025}
G. Tomassetti, “Direct and surrogate optimization in applied superconductivity: state of the art, perspectives and challenges,” \emph{Supercond. Sci. Technol.}, vol. 38, no. 7, p. 073001, 2025.

\bibitem{YazdaniAsrami2023}
M. Yazdani-Asrami \emph{et al.}, “Roadmap on artificial intelligence and big data techniques for superconductivity,” \emph{Supercond. Sci. Technol.}, vol. 36, no. 4, p. 043501, 2023.

\bibitem{Raissi2019}
M. Raissi, P. Perdikaris, and G. E. Karniadakis, “Physics-informed neural networks: A deep learning framework for solving forward and inverse problems involving nonlinear partial differential equations,” \emph{J. Comput. Phys.}, vol. 378, pp. 686-707, 2019.

\bibitem{AlipourBonab2025}
S. Alipour Bonab, W. Song, and M. Yazdani-Asrami, “Physics-informed neural network model for transient thermal analysis of superconductors,” \emph{Supercond. Sci. Technol.}, vol. 38, no. 8, p. 08LT01, 2025.

\bibitem{Cho2014}
K. Cho \emph{et al.}, “Learning phrase representations using RNN encoder-decoder for statistical machine translation,” \emph{arXiv:1406.1078}, 2014.

\bibitem{Leclerc2016}
J. Leclerc, L. M. Hell, C. Lorin, and P. J. Masson, “Artificial neural networks for AC losses prediction in superconducting round filaments,” \emph{Supercond. Sci. Technol.}, vol. 29, no. 6, p. 065008, 2016.

\bibitem{YazdaniAsrami2020} 
M. Yazdani-Asrami \emph{et al.}, “Prediction of nonsinusoidal AC loss of superconducting tapes using artificial intelligence-based models,” \emph{IEEE access}, vol. 8, pp. 207287–207297, 2020.

\bibitem{Zhou2025}
L. Zhou \emph{et al.}, “AC Loss Calculation of High Temperature Superconducting Coils Based on a Surrogate Model,” \emph{IEEE Trans. Appl. Supercond.}, vol. 35, no. 5, 2025, Art no. 5902205.

\bibitem{Purnode2024}
F. Purnode, F. Henrotte, G. Louppe, and C. Geuzaine, “Neural network-based simulation of fields and losses in electrical machines with ferromagnetic laminated cores,” \emph{International Journal of Numerical Modelling: Electronic Networks, Devices and Fields}, vol. 37, no. 2, p. e3226, 2024.

\bibitem{Kleeven2013}
W. Kleeven \emph{et al.}, “The IBA superconducting synchrocyclotron project S2C2,” in \emph{Proc. Cyclotrons}, 2013, pp. 115–119.

\bibitem{Carr2001}
W. J Carr Jr., \emph{AC loss and macroscopic theory of superconductors}. 2nd ed. London, UK: Taylor and Francis, 2001.

\bibitem{Dular1998}
P. Dular, C. Geuzaine, F. Henrotte, and W. Legros, “A general environment for the treatment of discrete problems and its application to the finite element method,” \emph{IEEE Trans. Magn.}, vol. 34, no. 5, pp. 3395-3398, 1998.

\bibitem{Vitrano2023}
A. Vitrano \emph{et al.}, “An Open-Source Finite Element Quench Simulation Tool for Superconducting Magnets,” \emph{IEEE Trans. Appl. Supercond.}, vol. 33, no. 5, 2023, Art no. 4702006.

\bibitem{Bottura2000}
L. Bottura, “A practical fit for the critical surface of NbTi,” \emph{IEEE Trans. Appl. Supercond.}, vol. 10, no. 1, pp. 1054-1057, 2000.

\bibitem{Spencer1979}
C. Spencer, P. Sanger, and M. Young, “The temperature and magnetic field dependence of superconducting critical current densities of multifilamentary Nb3Sn and NbTi composite wires”, \emph{IEEE Trans. Magn.}, vol. 15, no. 1, pp. 76-79, 1979.

\bibitem{Lubell1983}
M. S. Lubell, “Empirical scaling formulas for critical current and critical field for commercial NbTi”, \emph{IEEE Trans. Magn.}, vol. 19, no. 3, pp. 754-757, 1983.

\end{thebibliography}
\end{document}